\documentclass[preprintnumbers,amsmath,amssymb,floatfix,11pt,prd,onecolumn,showpacs,superscriptaddress,nofootinbib]{revtex4}
\usepackage{graphicx}
\usepackage{epsfig}
\usepackage{bm}
\usepackage{amsfonts}

\begin{document}
\title{Phase Space Analysis of Interacting Dark Energy in  $f(T)$ Cosmology}

\author{\textbf{ Mubasher Jamil}} \email{mjamil@camp.nust.edu.pk}
\affiliation{Center for Advanced Mathematics and Physics (CAMP),\\
National University of Sciences and Technology (NUST), H-12,
Islamabad, Pakistan} \affiliation{Eurasian International Center for
Theoretical Physics, Eurasian National University, Astana 010008,
Kazakhstan}

\author{\textbf{ Kuralay Yesmakhanova}}
 \affiliation{Eurasian International Center
for Theoretical Physics, Eurasian National University, Astana
010008, Kazakhstan}

\author{\textbf{ D. Momeni}}
\email{d.momeni@yahoo.com}
 \affiliation{Eurasian International Center
for Theoretical Physics, Eurasian National University, Astana
010008, Kazakhstan}

\author{\textbf{ Ratbay Myrzakulov}}
\email{rmyrzakulov@csufresno.edu}\affiliation{Eurasian International Center
for Theoretical Physics, Eurasian National University, Astana
010008, Kazakhstan}

\begin{abstract}
\textbf{Abstract:} In this paper, we
examine the interacting dark energy model in $f(T)$ cosmology.
We assume dark energy as a perfect fluid and choose a specific cosmologically viable form $f(T)=
\beta\sqrt{T} $. We show that there is one attractor solution to the dynamical equation of $f(T)$ Friedmann
equations. Further we investigate the stability in phase space for a
general $f(T)$ model with two interacting fluids.  By studying the
local stability near the critical points, we show that the critical
points lie on the sheet $u^*=(c-1)v^*$ in the phase space,
spanned by coordinates $(u,v,\Omega,T)$. From this critical sheet,
we conclude that the coupling between the dark energy and matter
$c\in (-2,0)$.

\end{abstract}

 \maketitle



\section{Introduction}

Astrophysical observations indicate that nearly seventy percent of
the cosmic energy density is hidden in some unknown `dark' sector
commonly called as `dark energy' (DE) \cite{perl,pad}. The remaining
contribution to the total energy density is contained in dark matter
and meager baryonic matter \cite{bachall}. Dark energy is described
phenomenologically by an equation of state (EoS) $p_{d}=w\rho_{d}$,
where $p_{d}$ and $\rho_d$ are the pressure and energy density of
dark energy, while $w$ is an EoS parameter which effectively
describes cosmic acceleration. To describe DE, one must have
$p_{d}<0$ and $w<0$. \textbf{General relativity (GR) offers only one
solution to this puzzle},
, namely `cosmological constant' which
suffers from fine tuning and coincidence problems. Since GR fails to
explain the cosmic accelerated expansion, one needs to modify
curvature or the matter part in the Einstein field equations. Some
notable examples are $f(R)$ gravity \cite{nojiri}, scalar-tensor
gravity \cite{jarv} and Lovelock gravity \cite{love} and more
recently $f(T)$ gravity \cite{f(T)}, to name a few. Other models
where matter action is modified include the bulk viscous stress
\cite{brevik} and the anisotropic stress \cite{wands} or some exotic
fluid like Chaplygin gas (CG) \cite{kamen}. \textbf{ One of the
crucial tests to check the viability of extended theories of gravity
is the potential detection of gravitational waves \cite{corda}.}

We assume a
phenomenological form of interaction between matter and dark energy
since these are the dominant components of the cosmic composition,
following \cite{lima}. The exact nature of this interaction is
beyond the scope of the paper and unresolved till we obtain a
consistent theory of quantum gravity. In these interacting dark
energy-dark matter models, the dark energy decays into matter at a
rate proportional to Hubble length. \textbf{The interacting dark energy scenario can successfully resolve the coincidence problem
then stable attractor solutions of the
Friedmann-Robertson-Walker (FRW) equations can be obtained \cite{quartin}.}\textbf{Some observational support to these models come
from the astrophysical observations \cite{wang22,orfeu}.} \textbf{For
motivation from field theory and particle physics of the
interacting dark energy, the interested reader is referred to
\cite{mac,sean}.}

The present paper is devoted to the study of dynamics of interacting
dark energy in $f(T)$ cosmology.  This theory is based on torsion
scalar $T$ rather than curvature scalar $R$. We assume the dark
energy in the form of a perfect fluid interacting with the matter.
\textbf{We study this model by the local stability method and than extend it for
dark energy satisfying a more general form of equation of state.}

\section{Basic equations}

\subsection{Basics of $f(T)$ gravity}

General relativity is a gauge theory of the gravitational field. It
is based on the equivalence principle. \textbf{However  it is not necessary
to work with Riemannian manifolds.} \textbf{There are some extended
theories such as Riemann-Cartan, in them the geometrical structure of the
theory has non-vanishing object of non-metricity.
} In these extensions, there are more
than one dynamical quantity (metric). For example this theory may be
constructed from the  metric, non-metricity and torsion
\cite{smalley}. \textbf{Ignoring from the non-metricity of the theory, we
can leave the Riemannian manifold and go to Weitznbock
spacetime, with torsion and zero local Riemann tensor.}
 One sample of
such theories  is called teleparallel gravity in which we are
working in a non-Riemannian manifold.\textbf{The dynamics of the metric is
determined using the scalar torsion $T$.}\textbf{ The fundamental quantities
in teleparallel theory are the vierbein (tetrad) basis vectors
$e^{i}_{\mu}$.} This basis is an orthogonal, coordinate free basis,
defined by the following equation
\begin{eqnarray}\nonumber
g_{\mu\nu}=e_{\mu}^{i}e_{\nu}^j \eta_{ij}.
\end{eqnarray}
This tetrad basis must be orthonormal and $\eta_{ij}$ is the
Minkowski metric. It means that $e^{i}_{\mu}e_{\mu j}=\delta^i_{j}$.
\textbf{There is a simple extension of the teleparallel gravity, which is
called  $f(T)$ gravity.} In this theories, f is an arbitrary function
of the torsion $T$. One suitable form of  action for $f(T)$ gravity
in  Weitzenbock spacetime is \cite{saridakis}
\begin{eqnarray}\nonumber
S=\frac{1}{2\kappa^2}\int d^4x e(T+f(T)+L_m).
\end{eqnarray}
Here $e=det(e^{i}_{\mu})$, $\kappa^2=8\pi G$. The dynamical quantity of the model is the scalar torsion $T$ and $L_m$ is the matter Lagrangian.
\textbf{The field equation can be derived from the action by varying the action
with respect to $e^{i}_{\mu}$}
\begin{eqnarray}\nonumber
&&e^{-1}\partial_{\mu}(e S^{\:\:\:\mu
\nu}_{i})(1+f_T)-e_i^{\:\lambda}T_{\:\:\:\mu
\lambda}^{\rho}S^{\:\:\:\nu \mu}_{\rho}f_T\nonumber\\&& +S^{\:\:\:\mu
\nu}_{i}\partial_{\mu}(T)f_{TT}-\frac{1}{4}e_{\:i}^{\nu}
(1+f(T))=4 \pi G e_i^{\:\rho}T_{\rho}^{\:\:\nu},
\end{eqnarray}
where $T_{\rho}^{\:\:\nu}$ is the energy-momentum tensor for matter sector of the Lagrangian $L_m$, defined by $$T_{\mu\nu}=-\frac{2}{\sqrt{-g}}\frac{\delta\int\Big(\sqrt{-g}L_m d^4x\Big)}{\delta g^{\mu\nu}}.$$ Here $T$ is defined by
\begin{equation}\nonumber
T=S^{\:\:\:\mu \nu}_{\rho} T_{\:\:\:\mu \nu}^{\rho},
\end{equation}
where
$$
T_{\:\:\:\mu \nu}^{\rho}=e_i^{\rho}(\partial_{\mu}
e^i_{\nu}-\partial_{\nu} e^i_{\mu}),
$$
$$
S^{\:\:\:\mu \nu}_{\rho}=\frac{1}{2}(K^{\mu
\nu}_{\:\:\:\:\:\rho}+\delta^{\mu}_{\rho} T^{\theta
\nu}_{\:\:\:\theta}-\delta^{\nu}_{\rho} T^{\theta
\mu}_{\:\:\:\theta}),
$$
and the contorsion tensor reads $K^{\mu \nu}_{\:\:\:\:\:\rho}$  as
$$
K^{\mu \nu}_{\:\:\:\:\:\rho}=-\frac{1}{2}(T^{\mu
\nu}_{\:\:\:\:\:\rho}-T^{\nu \mu}_{\:\:\:\:\:\rho}-T^{\:\:\:\mu
\nu}_{\rho}).
$$
\textbf{It is straightforward to show that this equation of motion reduces
to Einstein gravity when $f(T)=0$.} \textbf{Indeed, it is the equivalency
between the teleparallel theory and Einstein gravity \cite{T}.} The
theory has been found to address the issue of cosmic acceleration in
the early and late evolution of universe \cite{godel} but this
crucially depends on the choice of suitable $f(T)$, \textbf{for instance
exponential form containing $T$ cannot lead to phantom crossing
\cite{wuyu}}. Reconstruction of $f(T)$ models has been reported in
\cite{setare} \textbf{while thermodynamics of $f(T)$ cosmology including
the generalized second law of thermodynamics has been recently investigated
\cite{gsl}.}

\subsection{Interacting dark energy in $f(T)$ cosmology}

 We adopt the metric in the form of a flat Friedmann-Lemaitre-Robertson-Walker
  metric with  metric $ds^2=dt^2-a(t)^2(dx^idx_i)$, $i=1,2,3$.
\textbf{We start with the Friedmann equation for the $f(T)$ model \cite{f(T)}}
\begin{equation}\label{1-a}
H^2=\frac{1}{1+2f_T}\Big( \frac{\kappa^2}{3}\rho-\frac{f}{6} \Big),
\end{equation}
where $\rho=\rho_m+\rho_d$, and
$\rho_m$, $\rho_d$ represent the energy densities of
matter and dark energy.

\textbf{The second FRW equation is}
\begin{equation}\label{2-a}
\dot H=-\frac{\kappa^2}{2}\Big(\frac{\rho+p}{1+f_T+2Tf_{TT}}\Big).
\end{equation}
For a spatially flat universe ($k=0$),  the total energy
conservation equation is
\begin{equation}\label{1a}
\dot \rho+3H(\rho+p)=0,
\end{equation}
where $H$ is the Hubble parameter, $\rho$ is the total energy
density and $p$ is the total pressure of the background fluid.
The so-called energy-balance equations corresponding
to dark energy and dark matter are \cite{int}
\begin{eqnarray}\label{2a}
\dot \rho_d+3H(\rho_d+p_d)&=&-Q,\nonumber\\
\dot\rho_m+3H\rho_m&=&Q,
\end{eqnarray}
Here $Q$ is the interaction term which corresponds to energy
exchange between dark energy and dark matter.
The function $Q$ has dependencies on the energy
densities of the dark matter and dark energy and the Hubble parameter, i.e. $Q(H\rho_m)$,
$Q(H\rho_{d})$ or $Q(H\rho_{m},H\rho_{d})$ \cite{feng2}.
Since the nature of both dark energy and dark matter is unknown,
it is not possible to derive $\Gamma$ from first principles. To give a reasonable $Q$, we may expand like
$Q(H\rho_{m},H\rho_{d})\simeq \alpha_{m} H
\rho_{m}+\alpha_{d}H\rho_{d}$. Since the coupling strength is
also not known, we may adopt just one parameter for our convenience;
hence we take $\alpha_{m}=\alpha_{d}=c$ \cite{campo}. We here
choose the following coupling function $Q=3Hc(\rho_d+\rho_m)$.

To perform the stability analysis of the cosmological model, it is always convenient to define dimensionless density parameters via
\begin{equation}\label{3-a}
u\equiv\frac{\kappa^2\rho_d}{3H^2},\ \
v\equiv\frac{\kappa^2 p_d}{3H^2},\ \
\Omega_m\equiv\frac{\kappa^2\rho_m}{3H^2}.
\end{equation}
Moreover the equation of state parameter of dark energy
\begin{eqnarray}
w\equiv\frac{p_d}{\rho_d}=\frac{v}{u}.
\end{eqnarray}
Also the equation of state of all the fluids together is
\begin{eqnarray}\label{wtot}
w_{tot}\equiv\frac{p_d}{\rho_d+\rho_m}=\frac{v}{1-\Omega_T}.
\end{eqnarray}
Using (\ref{3-a}), we can rewrite (\ref{1-a}) in dimensionless form
\begin{eqnarray}
\Omega_m=1-u-\Omega_T,
\end{eqnarray}
where $\Omega_T\equiv\frac{f(T)}{T}-2f_T,$ is another dimensionless
density parameter constructed for torsion scalar. \textbf{In $f(T)$ and
teleparallel gravities, phase space analysis for different dark
energy models has been reported in \cite{sari,davood}.}

\section{Stability analysis of perfect fluid}

In this section, we treat dark energy as a perfect fluid. Physically it means that the fluid has no viscosity or heat transfer property. Its a simple fluid which cannot be self-gravitating and has a smooth distribution in space. A perfect fluid is represented by a phenomenological linear equation of state connecting energy density and pressure by
\begin{equation}\label{eos}
p_d=w \rho_d,
\end{equation}
where $w$ is a `constant' of proportionality but can be a function
of time or e-folding parameter $x=\ln a$. For a general dark energy
paradigm, the minimum condition to be satisfied is $w\leq-1/3$.
 \textbf{This opens a window to construct variety of theoretical models
 to explain cosmic acceleration.}

  Some of these well-known models are cosmological constant ($w=-1$), quintessence ($w<-1/3$)
  and phantom energy ($w<-1$).

The dynamical system representing the dynamics of perfect fluid's density and pressure reads as
\begin{eqnarray}
\frac{du}{dx}&=&-3[u+v+c(1-\Omega_T)]\nonumber\\&&+3u\Big[\frac{1-\Omega_T+v}{1+2Tf_T+f_{TT}}\Big],\label{sys1}\\
\frac{dv}{dx}&=&-3w[u+v+c(1-\Omega_T)]\nonumber\\&&+3v\Big[\frac{1-\Omega_T+v}{1+2Tf_T+f_{TT}}\Big].\label{sys2}
\end{eqnarray}
We notice that our dynamical system is under-determined: three
unknown functions ($u,v,\Omega_T$) and two coupled equations. Thus
we must choose $f(T)$ to solve the system. It is discussed in
\cite{bamba} that a power-law form of correction term $f(T)\sim T^n$
$ (n>1)$ such as $T^2$ can remove the finite-time future
singularity. However, when $n=0$, the correction term behaves like a
cosmological constant. \textbf{The model with $n=1/2$ can be helpful in
realizing power-law inflation, and also describes little-rip and
pseudo-rip cosmology \cite{bamba}.}
  Due to these reasons, we choose
$$f(T)= \beta\sqrt{-T}, $$ where $\beta$ is a constant. \textbf{Note that
choosing $\beta=0$ (or $f(T)=0$) leads to teleparallel gravity which
is equivalent to General Relativity. Note that this choice $f(T)=
\beta\sqrt{-T},$ has correspondence with the cosmological constant
EoS in $f(T)$ gravity \cite{mirza}.  This $f(T)$ model can be
recovered via reconstruction scheme of holographic dark energy
\cite{hde}. Also it can be inspired from a model for dark energy
model form the Veneziano ghost \cite{kk}. Recently Capozziello et al
\cite{capo123} investigated the cosmography of $f(T)$ cosmology by
using data of BAO, Supernovae Ia and WMAP. Following their
interesting results, we notice that if we choose
$\beta=\sqrt{3/2}H_0(\Omega_{m0}-1)$, than one can estimate the
parameters of this $f(T)$ model as a function of Hubble parameter
$H_0$ and the cosmographic parameters and the value of matter
density parameter. }

Recently attractor solutions for the dynamical system with three
fluids (dark matter, dark energy and radiation) interacting
non-gravitationally have been investigated to resolve the
coincidence problem using similar $f(T)$ \cite{davood}.

To perform the stability of the system comprising equations
(\ref{sys1}) and (\ref{sys2}), we first calculate the critical
points ($u_*,v_*$) by equating $$\frac{du}{dx}=0, \ \
\frac{dv}{dx}=0.$$ We find two critical points A and B for system
(\ref{sys1}) and (\ref{sys2}) given in Table-I. We check the
stability of the dynamical system in the neighborhood of these
critical points $u=u_{*}+\delta u, v=v_{*}+\delta v$. This is
performed by linearizing the system of equations of motion for $u$
and $v$ like
\begin{eqnarray}
{\frac {d\delta u}{dx}}  &= & \frac{3}{2}v_*\delta
u+ 3\left( -1+\frac{1}{2}u_* \right) \delta v,
\\
{\frac {d\delta v}{dx}}  &=&-3w\delta u+
3\left( 1-w +v_* \right) \delta v,
\end{eqnarray}
Next we construct a Jacobian matrix from the coefficients of $\delta
u$ and $\delta v$ in the linearized (or perturbed) system and
finding the eigenvalues from it. For the system (\ref{sys1}) and
(\ref{sys2}), the Jacobian matrix for any critical point is
\begin{eqnarray}
J=\left[ \begin {array}{cc} \frac{3}{2}v_*&-3+\frac{3}{2}u_*\\
\noalign{\medskip}-3 w&-3\,w+3+3\,v_*\end {array}
\right].
\end{eqnarray}

We find the eigenvalues $\lambda_{1,2}$ of the Jacobian matrix for
the two critical points given in Table-I. A critical point (also
called equilibrium point) is said to be \textit{stable} if the
corresponding eigenvalues are negative $\lambda_{1,2}<0$ for all the
values of the model parameters. Such stable critical points are
called
  attractor solutions of the dynamical system \textbf{i.e. solution of the dynamical system
  for different initial conditions which converge to a stable critical point.} A critical point is
   said to be \textit{unstable} if both the eigenvalues are positive, \textbf{and a \textit{saddle point}
    if one of the eigenvalues is positive.} We are interested only in stable critical points.
     Sometimes we get \textit{conditionally stable} critical points i.e.
      the critical point which can be stable only under some conditions on the model parameters.
      In cosmology conditionally stable points are also of interest. In Table-\ref{table:parameters}, we write down the values of relevant cosmological parameters including the total equation of state and the deceleration parameter. Note that for $f(T)=\beta\sqrt{T}$, we have $\Omega_T=0$, therefore $w_{tot}=v_*$. For A, $q<0$ when $c<1/3$.

\begin{table}
\centering
  \begin{tabular}{|c|c|c|c|}
  \hline\hline
    Point & $w_{tot}=v_*$ & $w=\frac{v_*}{u_*}$ & $q=-1-\frac{3}{2}(u_*+v_*)$ \\ [0.5ex]
     \hline
      A & $-2$ & $\frac{2}{c-1}$ &
$3c-1$ \\\hline
      B & $2\delta$ & $w$ & $1-3\delta(1+\frac{1}{w})$\\
      \hline
      \end{tabular}
        \caption{Cosmological implications.}
       \label{table:parameters}
       \end{table}

\begin{widetext}
  \begin{table}[ht]
  \centering
  \begin{tabular}{|c| c |c| c |c|}
  \hline\hline
    Point & $(u_*,v_*)$ & $\lambda_1$ & $\lambda_2$ & Stability  Condition \\ [0.5ex]
     \hline
      A & $(2(1-c),-2)$ &$3(-1-\frac{w}{2}\,+\frac{1}{2}\,\sqrt
{{w}^{2}+4\, w\,c})$ & $3(-1-\frac{w}{2}\,-\frac{1}{2}\,\sqrt
{{w}^{2} +4\,w\,c})$ & Unstable point \\\hline
      B & $( \frac{2\delta}{w}, 2\delta)$ &$3(1+\delta)$ & $3(\delta-w)$ & Stable point if
       $c<\frac{w+1}{w}$\\
      \hline
      \end{tabular}
        \caption{Critical points and stability of the perfect fluid model ( $\delta=\frac{1}{2}(w-\,\sqrt
{{w}^{2}+4\,w\,c})$, ${w}^{2}+4\,w\,c\geq0$ ).}
       \label{table:fluid}
       \end{table}
\end{widetext}

We observe that A is an unstable critical
point and B is conditionally stable. This means that if the
interaction parameter $c<\frac{w+1}{w}$, this point is a
stable critical point. Since our two dimensional phase space
embedded in a three dimensional space, \textbf{it has no chaotic behavior,
and we can investigate the dynamical behavior of the the system by
using the usual dynamical systems approach. }In figure 1, we plotted
the dynamical phase space of the (\ref{sys1}), (\ref{sys2}) for
different values of the parameters and for a range of the initial
conditions. We adopted the initial value (here initial value means the present value) for energy density of dark energy
$u(0)=0.7$, but different values of pressure parameter $v(0)$ \textbf{for different forms of dark energy in order to have a good agreement with the observational data.}\textbf{From figure-1 it turns out the trajectory of phantom energy in phase space} is more steep than quintessence and cosmological constant. From figure-2, we plot the deceleration parameter for different dark energy models in $f(T)$ cosmology. For cosmological constant, the trajectory starts near $x=-1$ (or $a=0.36$) and terminates close to $x=4$ (or $a=54.59$), meanwhile the deceleration parameter remains fixed between $x=-1$ and $x=4$ since Hubble parameter is constant. For quintessence, the deceleration parameter begins increasing from $x=-0.5$ ($a=0.6$) to  $x=1.5$ ($a=4.48$) where $q$ vanishes. Thus deceleration $q>0$ is not possible with quintessence dark energy. For phantom energy, $q$ starts decreasing from $x=-1.3$ but gets stable to value $q=-0.7$.

\begin{figure}
\centering
 \includegraphics[scale=0.3] {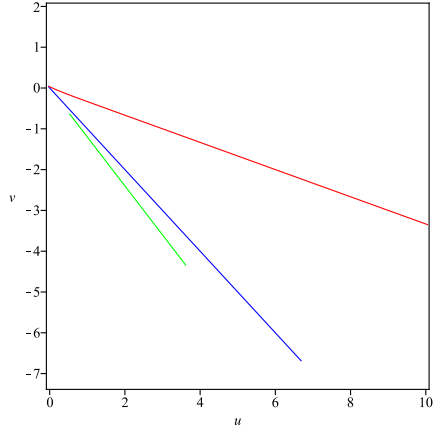}
  \caption{Phase space for perfect fluid form of dark energy. Three forms of dark energy are shown in figure: phantom energy $v(0)=-0.84$ ($w=-1.2$, green); cosmological constant $v(0)=-0.7$ ($w=-1$, blue); quintessence $v(0)=-0.231$ ($w=-0.33$, red).   }
  \label{1.eps}
\end{figure}

\begin{figure}
\centering
 \includegraphics[scale=0.3] {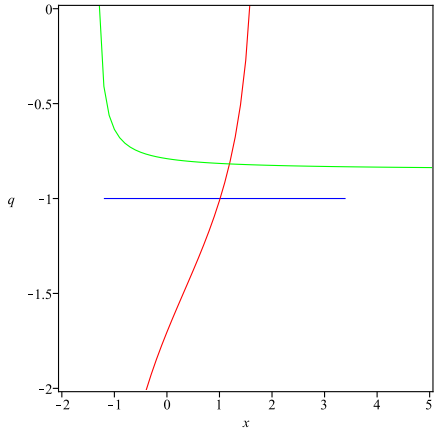}
  \caption{ Behavior of deceleration parameter for perfect fluid form of dark energy. Three forms of dark energy are shown in figure: phantom energy $v(0)=-0.84$ ($w=-1.2$, green); cosmological constant $v(0)=-0.7$ ($w=-1$, blue); quintessence $v(0)=-0.231$ ($w=-0.33$, red).
 }
  \label{2.eps}
\end{figure}

\section{The Phase space of the interacting dark energy models:
General case}

For a generic form of the $f(T)$, the system of
equations (\ref{sys1}), (\ref{sys2}) is not closed i.e. a third differential equation for $\Omega_T$ is needed.
 The EoS of the fluid is in the form $P_d=\Phi(\rho_d)$. The general dynamical
 system then reads
\begin{eqnarray}
\frac{du}{dx}&=&-3[c(1-\Omega_T)+u+v]\nonumber\\&&+\frac{3u(1-\Omega_T+v)}{1+f_T+2Tf_{TT}},\label{t1}\\
\frac{dv}{dx}&=&-3\Phi'[c(1-\Omega_T)+u+v]\nonumber\\&&+\frac{3v(1-\Omega_T+v)}{1+f_T+2Tf_{TT}},\label{t2}\\
\frac{d\Omega_T}{dx}&=&-\frac{3T(1-\Omega_T+v)(Tf_T-f-2T^2f_{TT})}{1+f_T+2Tf_{TT}}.\label{t3}
\end{eqnarray}
Here $\Phi'=\frac{d\Phi}{d\rho_d}.$ The system (\ref{t1}-\ref{t3}) is non-autonomous due to presence of $f$ and we need to add another equation to it. From the definition
of the $T=-6H^2$ it is easy to show that the fourth equation is
\begin{eqnarray}\label{t4}
\frac{dT}{dx}&=& -\frac{3T(1-\Omega _T+v)}{1+f_T+2Tf_{TT}}.
\end{eqnarray}
Now the system (\ref{t1}-\ref{t4}) is closed.
We discussed the different possible attractors of the system
(\ref{t1}-\ref{t4}) in Table-\ref{table:f(T)}.

 \begin{table}[ht]
  \caption{Critical Points of the  general $f(T)$ model.}
  \centering
  \begin{tabular}{|c| c| c|}
  \hline\hline
    Critical sheet & $(u^*,v^*,\Omega_T^*,T^*)$ & Stability  Condition \\ [0.5ex]
     \hline
      C & $(u^*,v^*,\Omega_T^*(0),0)$  &  Physically unacceptable \\
      D & $( (c-1)v^*,v^*,T^*,,\Omega_T^*)$ & Conditionally stable \\
      \hline
      \end{tabular}
       \label{table:f(T)}
       \end{table}

Case C is not physically viable since when  $T^* = 0$ it
implies $H=0$. It means that the local geometry in the
\textbf{neighbourhood} of this point is Minkowski flat. We are not
interested to such cases since astrophysically $H\neq0$. But for case D, we conclude that the
critical points in the phase space lie on the two
dimensional surface
\begin{eqnarray}
u^*=(c-1)v^*.
\end{eqnarray}
Since EoS
is $w=\frac{v}{u}$, this equation tells us that the critical point
can be written in the form $p_d^*=\frac{\rho_d^*}{c-1}$. It shows
that $w^*=\frac{1}{c-1}$. Since $-1<w^*<-\frac{1}{3}$, it shows that the interacting coupling must be in the range
$c\in(-2,0)$, \textbf{which is also reported in an earlier work \cite{eur}.} This is a new constraint on the coupling constant $c$
from the phase analysis approach. But from this analysis we can not
obtain any new information about the values of the $T^*,\Omega_T^*$.
We remember here that, this discussion is free from any
specific form of the $f(T)$. We conclude that for interacting dark
energy models in $f(T)$ gravity, there is only one physically
acceptable critical point D.

\section{Conclusion}

\textbf{In this paper, we investigated the stability and phase space
description of a perfect fluid form of the dark energy interacting
with matter.} We wrote the general dynamical system equations for two
fluids and found the critical points.\textbf{ We linearized the system of
equations and showed that for perfect fluid case there is only
one attractor solution.}
 For any value of the EoS parameter $w$, we
obtained a new bound on the coupling constant $c<\frac{w+1}{w}$. It
is a new constraint on $c$ in the context of the $f(T)$ gravity
which shows that energy is being transferred from matter to dark
energy. Thus we obtain a scenario where the universe becomes
increasingly dark energy dominated. Further for a general $f(T)$
model, there is only one critical point, and this critical point
\textbf{lives on the surface with equation $u^*=(c-1)v^*$ in the four
dimensional phase space which is spanned by four coordinates
$X=(u^*,v^*,\Omega_T^*,T^*)$.}

\subsection*{Acknowledgment}
\textbf{The authors would like to thank the anonymous referees for
their enlightening comments on our paper for its improvement.}

\end{document}